May 30, 2008

# A brief comment on the chemical formulae of the rare earth iron arsenide oxide superconductors

Z. Hiroi

*Institute for Solid State Physics, University of Tokyo, Kashiwa, Chiba 277-8581, Japan*

The recent discovery of superconductivity at rather high critical temperatures in a lanthanum iron arsenide oxide doped with fluorine by Kamihara, Watanabe, Hirano and Hosono[1] has triggered a "fever" in superconductivity research, just as happened about 20 years ago for copper oxide superconductors[2] or 7 years ago for $MgB_2$.[3] This is obviously because of its high $T_c$ values, 26 K at ambient pressure[1] and 43 K under high pressures.[4] Moreover, higher $T_c$'s (over 50 K) have been reported in other rare earth compounds,[5,6] tending to renew our optimism towards higher $T_c$'s, as with the copper oxides and their successors. In addition, a possible exotic mechanism, based on a magnetic fluctuation on the square lattice of the iron atoms, is being widely discussed and is stimulating much research into their physical properties.[7-9] In contrast to previous feverish research activity, however, is the surprisingly rapid acceleration of research. More than several papers have already been published in journals within four months of the first report[4-12] and so many papers have been uploaded on to the preprint server arXiv.org that it is impossible for this comment to cover everything. This may be because everybody has been hoping for a new superconductor with a high $T_c$ to appear and is ready with quick experiments and theories. Of course, the development of the preprint server is itself another factor in this rapid growth, as well as being somewhat anarchical.

In this turbulent situation, I have a particular concern about the chemical formulae of this family of compounds, from a chemistry point of view. There have been two kinds of formulae used so far in the literature. One is $LaFeAsO$ or $LaFeAsO_{1-x}F_x$/$LaFeAsO_{1-\delta}$ when it is rendered superconducting[5,7,11-13] (Here I use La as representing one of the rare earth elements). The other is $LaOFeAs$ or $LaO_{1-x}F_x FeAs$/$LaO_{1-\delta}FeAs$.[1,4,6,8-10] To call one compound by different names is no doubt unfavorable and must lead to serious confusion in the future.

Generally, the sequence of citation of symbols in formulae is decided according to the rules provided by the International Union of Pure and Applied Chemistry (IUPAC), which may be found partly on the web and references therein (http://old.iupac.org/reports/provisional/abstract04/connelly_310804.html) or in more detail in the book "Nomenclature of Inorganic Chemistry, IUPAC Recommendations 2005".[14] For an inorganic compound, the basic idea for a generalised salt is to take electronegativity as the ordering principle in its formula. The atomic symbols are cited according to their relative electronegativities, the least electronegative element being cited first and the most electronegative one last.  The exception is a class of coordination compounds or chain compounds which contain a well-defined polyatomic group or a linear sequence of atoms in the structure.  Since the present compound is definitely a generalised salt, the electronegativity rule must be applied to decide the formula.   Since the electronegativities are 1.1, 1.83, 2.18 and 3.44 for La, Fe, As, and O, respectively, the chemical formula is uniquely determined to be LaFeAsO and not LaOFeAs.

The chemical formula of an inorganic compound should not depend critically on its structure, unless it contains polyatomic groups such as $NH_4$, $SO_4$, etc. A layered compound is no exception. A good example is found in the copper oxide superconductors that usually contain a $CuO_2$ layer as a well-defined structural and also electronic unit.  The most famous one is $La_2CuO_4$ that crystallizes in a layer sequence: $La_2O_2/CuO_2$.  However, nobody calls it "$La_2O_2CuO_2$".  Another good example is $Ca_2CuO_2Cl_2$, not "$Ca_2Cl_2CuO_2$", although it has a layered structure made up of $Ca_2Cl_2$ and $CuO_2$ layers.[15]   Therefore, it is not rational to name "LaOFeAs" from the structural viewpoint, even if one considers that it is a layered compound consisting of LaO and FeAs layers.

Personally, I do not in any case agree that this compound should be called a layered compound from the viewpoint of crystal chemistry.  Generally speaking, a layered compound should have a structure where rigid layers with strong chemical bonding within the layers stack via a weak van der Waals coupling.   In the crystal structure of LaFeAsO, it is only marginal to assume a layer stacking of LaO and FeAs, because the La atom is in fact coordinated by 4 oxide atoms and 4 As atoms; a La cation always prefers to be surrounded by many more than six anions.  Moreover, band structure calculations have found that most electronic states near the Fermi level come from Fe $3d$ states,[8] so that it is more realistic to assume that a conductive Fe layer is sandwiched by insulating LaAsO layers.  Thus, it is not a conventional layered compound, although it is a quasi-two-dimensional system in terms of its electronic structure. Again, there is no reason to use the name "LaOFeAs".

The correct formula LaFeAsO is, in fact, the one used by Quebe, Terbüchte and Jeitschko in their first paper reporting the discovery of the series of quaternary compounds $RT$AsO ($R$: La, Ce, Pr, Nd, Sm, Gd; $T$: Fe, Ru, Co) with the ZrCuSiAs type structure.[13]  The subsequent discovery of superconductivity by doping one of these parent compounds with fluorine by Kamihara *et al*. is, of course, a great advance.[1]   However, one should also respect the original work by those distinguished chemists.   Since this family of compounds will, no doubt, turn out to be one of the most important systems in the field of solid state physics, I hope that everybody interested will use only the correct chemical formulae in their work, such as *RT*AsO, *RT*AsO$_{1-x}$F$_x$. (with F placed last, because it is the most electronegative ion), or *RT*AsO$_{1-\delta}$.

Let us enjoy the interesting chemistry and physics of these superconductors!